\documentclass{amsart}

\usepackage{amsmath}
\usepackage{amsthm}
\usepackage{amsfonts}
\usepackage{amssymb}
\usepackage{graphicx}
\usepackage{color}
\usepackage{ucs}
\usepackage[utf8x]{inputenc}

\definecolor{myurlcolor}{rgb}{0,0,0.7}
\usepackage[bookmarks=false]{hyperref}
\hypersetup{colorlinks,
linkcolor=myurlcolor,
citecolor=myurlcolor,
urlcolor=myurlcolor}

\newcommand{\gt}{>}

\newcommand{\maps}{\colon}
\newcommand{\define}[1]{{\bf{\boldmath{#1}}}}
\newcommand{\R}{\mathbb{R}}
\newcommand{\C}{\mathbb{C}}

\title{Quantropy} 
\date{\today}

\author{John C. Baez}
\address{Department of Mathematics\\ 
University of California\\ 
Riverside CA 92521\\
and Centre for Quantum Technologies\\ 
National University of Singapore\\ 
Singapore 117543}
\email{baez@math.ucr.edu}

\author{Blake S.\ Pollard} 
\address{Department of Physics\\ 
University of California\\ 
Riverside CA 92521 }
\email{bpoll002@ucr.edu}

\begin{document}

\maketitle

\begin{abstract}
\noindent
There is a well-known analogy between statistical and quantum mechanics.  In statistical mechanics, Boltzmann realized that the probability for a system in thermal equilibrium to occupy a given state is proportional to $\exp(-E/kT)$ where $E$ is the energy of that state.  In quantum mechanics, Feynman realized that the amplitude for a system to undergo a given history is proportional to $\exp(-S/i\hbar)$ where $S$ is the action of that history.   In statistical mechanics we can recover Boltzmann's formula by maximizing entropy subject to a constraint on the expected energy.  This raises the question: what is the quantum mechanical analogue of entropy?   We give a formula for this quantity, which we call `quantropy'.   We recover Feynman's formula from assuming that histories have complex amplitudes, that these amplitudes sum to one, and that the amplitudes give a stationary point of quantropy subject to a constraint on the expected action.  Alternatively, we can assume the amplitudes sum to one and that they give a stationary point of a quantity we call `free action', which is analogous to free energy in statistical mechanics.  We compute the quantropy, expected action and free action for a free particle, and draw some conclusions from the results.
\end{abstract}

\section{Introduction}

There is a famous analogy between statistical mechanics and quantum mechanics.   In statistical mechanics, a system can be in any state, but its probability of being in a state with energy $E$ is proportional to $\exp(-E/T)$ where $T$ is the temperature in units where Boltzmann's constant is 1.  In quantum mechanics, a system can move along any path, but its amplitude for moving along a path with action $S$ is proportional to $\exp(- S/i\hbar)$ where $ \hbar$ is Planck's constant.  So, we have an analogy where making the replacements
 \[ \begin{array}{rcl} E &\mapsto& S \\
 T &\mapsto& i \hbar \end{array}\] 
formally turns the probabilities for states in statistical mechanics into the amplitudes for paths, or `histories', in quantum mechanics.  In statistical mechanics, the strength of thermal fluctuations is governed by $T$.  In quantum mechanics, the strength of quantum fluctuations is governed by $\hbar$.

In statistical mechanics, the probabilities $ \exp(-E/T)$ arise naturally from maximizing entropy subject to a constraint on the expected value of energy. Following the analogy, we might guess that the amplitudes $ \exp(-S/i\hbar)$ arise from maximizing some quantity subject to a constraint on the expected value of action.  This quantity deserves a name, so let us tentatively call it `quantropy'.  

In fact, Lisi \cite{Lisi} and Munkhammar \cite{Munkhammar} have already treated quantum systems as interacting with a `heat bath' of action and sought to derive quantum mechanics from a principle of maximum entropy with amplitudes---or as they prefer to put it, complex probabilities---replacing probabilities.   However, seeking to derive amplitudes for paths in quantum mechanics from a maximum principle is not quite correct.  Quantum mechanics is rife with complex numbers, and it makes no sense to maximize a complex function.  But a complex function can still have stationary points, where its first derivative vanishes.   So, a less naive program is to derive the amplitudes in quantum mechanics from a `principle of stationary quantropy'.  We do this for a class of discrete systems, and then illustrate the idea with the example of a free particle, discretizing both space and time.

Carrying this out rigorously is not completely trivial.   In the simplest case, entropy is defined as a sum involving logarithms.  Moving to quantropy, each term in the sum
involves a logarithm of complex number.  Making each term well-defined requires a choice of branch cut; it is not immediately clear that we can do this and obtain
a differentiable function as the result.  Additional complications arise when we 
consider the continuum limit of the free particle.  Our treatment handles all these issues.

We begin by reviewing the main variational principles in physics and pointing out the conceptual gap that quantropy fills.  In Section \ref{quantropy} we introduce quantropy
along with two related quantities: the free action and the expected action.  In Section \ref{computing} we develop tools for computing all these quantities.  In Section \ref{free_particle} we illustrate our methods with the example of a free particle, 
and address some of the conceptual questions raised by our results.  We conclude by mentioning some open issues in Section \ref{conclusions}.


\subsection{Statics}

Static systems at temperature zero obey the \emph{principle of minimum energy}. In classical mechanics, energy is often the sum of kinetic and potential energy:
\[ E = K + V \]
where the potential energy $V$ depends only on the system's position, while the kinetic energy $K$ also depends on its velocity.  Often, though not always, the kinetic energy has a minimum at velocity zero.  In classical mechanics this lets us minimize energy in a two-step way.  First we minimize $K$ by setting the velocity to zero.  Then we minimize $V$ as a function of position.  

While familiar, this is actually somewhat noteworthy.  Usually minimizing the sum of two things involves an interesting tradeoff.   In quantum physics, a tradeoff really is required, thanks to the uncertainty principle.  We cannot know the position and velocity of a particle simultaneously, so we cannot simultaneously minimize potential and kinetic energy.  This makes minimizing their sum much more interesting.  But in classical mechanics, in situations where $K$ has a minimum at velocity zero statics at temperature zero is governed by a \emph{principle of minimum potential energy}.  

The study of static systems at nonzero temperature deserves to be called `thermostatics', though it is usually called `equilibrium thermodynamics'.  In classical or quantum equilibrium thermodynamics at any fixed temperature, a system is governed by the \emph{principle of minimum free energy}.   Instead of our system occupying a single definite state, it will have different probabilities of occupying different states, and these
probabilities will be chosen to minimize the free energy
\[ F = \langle E \rangle - T S.\]
Here $\langle E \rangle$ is the expected energy, $T$ is the temperature, and $S$ is the entropy.    Note that the principle of minimum free energy reduces to the principle of minimum energy when $T = 0$.  

But where does the principle of minimum free energy come from?  One answer is that free energy $F$ is the amount of `useful' energy: the expected energy $\langle E \rangle$ minus the amount in the form of heat, $T S$.  For some reason, systems in equilibrium minimize this.  

Boltzmann and Gibbs gave a deeper answer in terms of entropy.  Suppose that our system has some space of states $X$ and the energy of the state $x \in X$ is $ E(x)$.   Suppose that $X$ is a measure space with some measure $dx$, and assume that we can describe the equilibrium state using a probability distribution, a function $p \maps X \to [0,\infty)$ with
\[           \int_X p(x) \, dx = 1 . \]
Then the entropy is
\[ \displaystyle{ S = - \int_X p(x) \ln p(x) \, dx}.\]
while the expected value of the energy is:
 \[  \displaystyle{ \langle E \rangle = \int_X E(x) p(x) } \, dx.\]
Now suppose our system maximizes entropy subject to a constraint on the expected value of energy.   Using the method of Lagrange multipliers, this is the same as maximizing $ S - \beta \langle E \rangle $ where $ \beta$ is a Lagrange multiplier.  When we maximize this, we see the system chooses a Boltzmann distribution:
\[ \displaystyle{ p(x) = \frac{\exp(-\beta E(x))}{\int_X \exp(-\beta  E(x))}}.\]
One could call $ \beta$ the \define{coolness}, since working in units where Boltzmann's constant equals 1 it is just the reciprocal of the temperature.   So, when the temperature is positive, maximizing $ S - \beta \langle E\rangle $ is the same as minimizing the free energy:
\[ F = \langle E \rangle - T S. \]

In summary: every minimum or maximum principle in statics can be seen as a special case or limiting case of the \emph{principle of maximum entropy}, as long as we admit that sometimes we need to maximize entropy subject to constraints.  This is quite satisfying, because as noted by Jaynes, the principle of maximum entropy is a general principle for reasoning in situations of partial ignorance \cite{Jaynes}.  So, we have a kind of `logical' explanation for the laws of statics. 


\subsection{Dynamics}

Now suppose things are changing as time passes, so we are doing dynamics instead of statics.  In classical mechanics we can imagine a system tracing out a path $q(t)$ as time passes from $t = t_0$ to $ t = t_1.$   The action of this path is often the integral of the kinetic minus potential energy:
\[ A(q) = \displaystyle{ \int_{t_0}^{t_1}  (K(t) - V(t)) \, dt }\]
where $ K(t)$ and $ V(t)$ depend on the path $q.$  To keep things from getting any more confusing than necessary, we are calling action $A$ instead of the more usual $ S,$ since we are already using $S$ for entropy.

The \emph{principle of least action} says that if we fix the endpoints of this path, that is the points $q(t_0)$ and $q(t_1),$ the system will follow the path that minimizes the action subject to these constraints.  This is a powerful idea in classical mechanics.  But in fact, sometimes the system merely chooses a stationary point of the action.  The Euler--Lagrange equations can be derived just from this assumption.  So, it is better to speak of the \emph{principle of stationary action}.

This principle governs classical dynamics.  To generalize it to quantum dynamics, Feynman proposed that instead of our system following a single definite path, it can follow any path, with an amplitude $a(q)$ of following the path $q.$  He proposed this formula for the amplitude:
\[ \displaystyle{ a(q) = \frac{\exp(i A(q)/\hbar)}{ \int  \exp(i A(q)/\hbar) \, dq}} \]
where $ \hbar$ is Planck's constant.  He also gave a heuristic argument showing that as $ \hbar \to 0$, this prescription reduces to the principle of stationary action.

Unfortunately the integral over all paths is hard to make rigorous except in certain special cases.   This is a bit of a distraction for our discussion now, so let us talk more abstractly about `histories' instead of paths with fixed endpoints, and consider a system whose possible histories form some space $X$ with a measure $dx$.   We will look at an example later.

Suppose the action of the history $ x \in X$ is $ A(x).$  Then Feynman's sum over histories formulation of quantum mechanics says the amplitude of the history $ x$ is: 
\[ \displaystyle{ a(x) = \frac{\exp(i A(x) /\hbar)}{\int_X  \exp(i A(x) /\hbar) }}. \]
This looks very much like the Boltzmann distribution:
\[ \displaystyle{  p(x) = \frac{\exp(- E(x)/T)}{\int_X \exp(-  E(x)/T)}}. \]
Indeed, the only serious difference is that we are taking the exponential of an imaginary quantity instead of a real one.  This suggests deriving Feynman's formula from a stationary principle, just as we can derive the Boltzmann distribution by maximing entropy subject to a constraint.  This is where quantropy enters the picture.


\section{Quantropy}
\label{quantropy}

We have described statics and dynamics, and a well-known analogy between them.  However, we have seen there are some missing items in the analogy:

\vskip 1em
\begin{center}
\renewcommand{\arraystretch}{1.5}
\begin{tabular}[h!]{|c| c|}
\hline
\textbf{Statics} & \textbf{Dynamics} \\
\hline
statistical mechanics & quantum mechanics\\
\hline
probabilities & amplitudes \\
\hline
Boltzmann distribution  & Feynman sum over histories \\
\hline
energy & action\\
\hline
temperature & Planck's constant times $i$  \\
\hline
entropy & ??? \\
 \hline
free energy & ??? \\
\hline
\end{tabular}

\end{center}
\vskip 1em

Our goal now is to fill in the missing entries in this chart.  Since the Boltzmann distribution 
\[ \displaystyle{  p(x) = \frac{\exp(- E(x)/T)}{\int_X \exp(-  E(x)/T) \, dx}} \]
 comes from the principle of maximum entropy, one might hope Feynman's sum over histories formulation of quantum mechanics:
\[ \displaystyle{ a(x) = \frac{\exp(i A(x) /\hbar)}{\int_X  \exp(i A(x) /\hbar) \, dx}} \] 
comes from a maximum principle as well.

Unfortunately Feynman's sum over histories involves complex numbers, and it does not make sense to maximize a complex function.  So let us try to derive Feynman's prescription from a \emph{principle of stationary quantropy}.

Suppose we have a set of histories, $X,$ equipped with a measure $dx$.  Suppose there is a function $a \maps X \to \C$ assigning to each history $ x \in X$ a complex amplitude $a(x)$.  We assume these amplitudes are normalized so that 
\[ \int_X a(x) \, dx= 1, \]
since that is what Feynman's normalization actually achieves. 
 We define the \define{quantropy} of $a$ by:
\[ \displaystyle{ Q = - \int_X a(x) \ln a(x) \, dx}.\]
One might fear this is ill-defined when $ a(x) = 0,$ but that is not the worst problem; in the study of entropy we typically set $ 0 \ln 0 = 0$. The more important problem is that the logarithm has different branches: we can add any multiple of $ 2 \pi i$ to our logarithm and get another equally good logarithm.  For now suppose we have chosen a specific logarithm for each number $ a(x),$ and suppose that when we vary the numbers $a(x)$ they do not go through zero.  This allows us to smoothly change $\ln a(x)$ as a function of $a(x)$.

To formalize this we could treat quantropy as depending not on the amplitudes $a(x)$, but on some function $b \maps X \to \C$ such that $\exp(b(x)) = a(x)$.   In this approach we require
\[ \int_X e^{b(x)} \, dx= 1, \]
and define the quantropy by:
\[ \displaystyle{ Q = - \int_X e^{b(x)} \, b(x) \, dx}.\]
Then the problem of choosing branches for the logarithm does not come up.  But we shall take the informal approach where we express quantropy in terms of amplitudes and choose a branch for $\ln a(x)$ as described above.

Next, let us seek amplitudes $ a(x)$ that give a stationary point of the quantropy $ Q$ subject to a constraint on the \define{expected action}:
\[  \displaystyle{ \langle A \rangle = \int_X A(x) a(x) \, dx } .\]
The term `expected action' is a bit odd, since the numbers $a(x)$ are amplitudes rather than probabilities.   While one could try to justify this term from how expected values are computed in Feynman's formalism, we are mainly using it because $\langle A \rangle$ is analogous to the expected value of the energy, $\langle E \rangle$, which we saw earlier.

Let us look for a stationary point of $Q$ subject to a constraint on $\langle A \rangle$, say $\langle A \rangle = \alpha$.  To do this, one would be inclined to use Lagrange multipliers and look for a stationary point of 
\[ Q - \lambda \langle A\rangle.\]
But there is another constraint too, namely
\[ \int_X a(x) \, dx = 1. \]
So let us write
\[ \langle 1 \rangle = \int_X a(x) \, dx \]
and look for stationary points of $ Q$ subject to the constraints 
\[ \langle A \rangle = \alpha  ,  \qquad \langle 1 \rangle = 1.\] 
To do this, the Lagrange multiplier recipe says we should find stationary points of 
\[ Q - \lambda \langle A \rangle - \mu \langle 1 \rangle \] 
where $ \lambda$ and $ \mu$ are Lagrange multipliers.  The Lagrange multiplier $ \lambda$ is the more interesting one.  It is analogous to the `coolness' $ \beta = 1/T,$ so our analogy chart suggests that we should take
\[ \lambda = 1/i\hbar.\] 
We shall see that this is correct.  When $\lambda$ becomes large our system becomes close to classical, so we call $ \lambda$ the \define{classicality} of our system. 

Following the usual Lagrange multiplier recipe, we seek amplitudes for which
\[  \displaystyle{ \frac{\partial}{\partial a(x)} \left(Q - \lambda \langle A\rangle - \mu \langle 1 \rangle \right) }  = 0\]
 holds, along with the constraint equations.  We begin by computing the derivatives we need: 
\[ \begin{array}{ccl} \displaystyle{ \frac{\partial Q}{\partial a(x)}  }  
&=& -(1 + \ln a(x) ) \\    \\    \displaystyle{ \frac{\partial \langle A \rangle}{\partial a(x)}  }  
&=& A(x) \\    \\  
 \displaystyle{ \frac{\partial \langle 1 \rangle}{\partial a(x)}   }  
&=& 1. \end{array} \]
Thus, we need 
\[  1+ \ln a(x) + \lambda A(x) + \mu  = 0 \]
 or
\[ \displaystyle{ a(x) = \frac{\exp(-\lambda A(x))}{\exp(\mu + 1)} }. \]
The constraint 
\[ \int_X a(x) \, dx= 1\] 
then forces us to choose
\[ \displaystyle{ \exp(\mu + 1) = \int_X \exp(-\lambda A(x)) \,dx} \]
so we have 
\[ \displaystyle{ a(x) = \frac{\exp(-\lambda A(x))}{\int_X \exp(-\lambda A(x)) \, dx }}. \]
This is precisely Feynman's sum over histories formulation of quantum mechanics if $\lambda = 1/i\hbar$!   

Note that the final answer does two equivalent things in one blow:
\begin{itemize}
\item It gives a stationary point of quantropy subject to the constraints that the amplitudes sum to 1 and the expected action takes some fixed value.  
\vskip 1em
\item It gives a stationary point of the \define{free action}:
\[  \langle A \rangle - i \hbar Q \]
subject to the constraint that the amplitudes sum to 1.
\end{itemize}
In case the second point is puzzling, note that the `free action' plays the same role in quantum mechanics that the free energy $ \langle E \rangle - T S$ plays in statistical mechanics.   It completes the analogy chart at the beginning of this section.  It is widely used in the effective action approach to quantum field theory, though not under the name `free action': as we shall see, it is simply $-i \hbar$ times the logarithm of the partition function. 

It is also worth noting that when $ \hbar \to 0$, the free action reduces to the action.  Thus, in this limit, the principle of stationary free action reduces to the principle of stationary action in classical dynamics.


\section{Computing Quantropy}
\label{computing}

In thermodynamics there is a standard way to compute the entropy of a system in equilibrium starting from its partition function.  We can use the same techniques to compute quantropy. It is harder to get the integrals to converge in interesting examples.  But we can worry about that later, when we do an example.  

First recall how to compute the entropy of a system in equilibrium starting from its partition function.  Let $X$ be the set of states of the system.  We assume that $X$ is a measure space, and that the system is in a mixed state given by some probability distribution $p \maps X \to [0,\infty)$, where of course 
\[ \int_X p(x) \, dx = 1 .\]
We assume each state $x$ has some energy $E(x) \in \R$.  Then the mixed state
maximizing the entropy
\[ S = - \int_X p(x) \ln p(x) \, dx \]
with a constraint on the expected energy
\[ \langle E \rangle = \int_X E(x) p(x) \, dx \]
is the Boltzmann distribution
\[ \displaystyle{ p(x) = \frac{e^{-\beta E(x)}}{Z} } \]
for some value of the coolness $\beta$, where $Z$ is the partition function:
\[ Z = \int_X e^{-\beta E(x)} \, dx .\]
To compute the entropy of the Boltzmann distribution, we can thus take the formula for entropy and substitute the Boltzmann distribution for $p(x)$, getting
\[ \begin{array}{ccl} S &=& \displaystyle{ 
\int_X p(x) \left( \beta E(x) + \ln Z \right)\, dx} \\   \\
&=& \beta \, \langle E \rangle + \ln Z .\end{array}  \]
Reshuffling this, we obtain a formula for the free energy:
\[   F = \langle E \rangle - T S = - T \ln Z .\]

Of course, we can also write the free energy in terms of the partition function and $\beta$:
\[ \displaystyle{ F = - \frac{1}{\beta} \ln Z }. \]
We can do the same for the expected energy:
\[ \begin{array}{ccl} \langle E \rangle &=& \displaystyle{ \int_X E(x) p(x) \, dx } \\   \\
&=& \displaystyle{ \frac{1}{Z} \int_X E(x) e^{-\beta E(x)} \, dx } \\   \\
&=& \displaystyle{ -\frac{1}{Z} \frac{d}{d \beta} \int_X e^{-\beta E(x)} \, dx } \\ \\
&=& \displaystyle{ -\frac{1}{Z} \frac{dZ}{d \beta} } \\ \\
&=& \displaystyle{ - \frac{d}{d \beta} \ln Z } .\end{array} \]
This in turn gives
\[ \begin{array}{ccl} S &=& \beta \, \langle E \rangle + \ln Z \\ \\ &=& \displaystyle{ - \beta \, \frac{d \ln Z}{d \beta} + \ln Z }.
\end{array} \]
In short: if we know the partition function of a system in thermal equilibrium as a function of $\beta$, we can easily compute its entropy,  expected energy and free energy.  

Similarly, if we know the partition function of a quantum system as a function of 
$\lambda = 1/i \hbar$, we can compute its quantropy, expected action and free action.
Let $X$ be the set of histories of some system.  We assume that $X$ is a measure space, and that the amplitudes for histories are given by a function
$a \maps X \to \C$ obeying
\[ \int_X a(x) \, dx = 1 .\]
We also assume each history $x$ has some action $A(x) \in \R$.  In the last section,
we saw that to obtain a stationary point of quantropy
\[ Q = - \int_X a(x) \ln a(x) \, dx \]
with a constraint on the expected action
\[ \langle A \rangle = \int_X A(x) a(x) \, dx \]
we must use Feynman's prescription for the amplitudes:
\[ \displaystyle{ a(x) = \frac{e^{-\lambda A(x)}}{Z} } \]
for some value of the classicality $\lambda = 1/i\hbar$, where $Z$ is the partition function:
\[ Z = \int_X e^{-\lambda A(x)} \, dx .\]

As mentioned, the formula for quantropy here is a bit dangerous, since we are taking the logarithm of the complex-valued function $a(x)$, which requires choosing a branch. Luckily, the ambiguity is greatly reduced when we use Feynman's prescription for $a$, because in this case $ a(x)$ is defined in terms of an exponential.  So, we can choose this branch of the logarithm:
\[ \ln a(x) = \displaystyle{ \ln \left( \frac{e^{iA(x)/\hbar}}{Z}\right) } = \frac{i}{\hbar} A(x) - \ln Z . \]
Once we choose a logarithm for $Z$, this formula defines $ \ln a(x)$.

Inserting this formula for $\ln a(x)$ into the formula for quantropy, we obtain
\[ \displaystyle{ Q = - \int_X a(x) \left( \frac{i}{\hbar} A(x) - \ln Z \right)\, dx } \]
We can simplify this a bit, since the integral of $a$ is 1:
\[ \displaystyle{ Q = \frac{1}{i \hbar} \langle A \rangle + \ln Z } . \]
We thus obtain:
\[  - i \hbar \ln Z = \langle A \rangle - i \hbar Q. \]
This quantity is what we called the `free action' in the previous section.  Let us denote it by the letter $\Phi$:
\[ \Phi = - i \hbar \ln Z. \]

In terms of $\lambda$, we have
\[ \displaystyle{ a(x) = \frac{e^{- \lambda A(x)}}{Z} }. \]
Now we can compute the expected action just as we computed the expected energy in thermodynamics:
\[ \begin{array}{ccl} \langle A \rangle &=& \displaystyle{ \int_X A(x) a(x) \, dx } \\ \\
&=& \displaystyle{ \frac{1}{Z} \int_X A(x) e^{-\lambda A(x)} \, dx } \\   \\
&=& \displaystyle{ -\frac{1}{Z} \frac{d}{d \lambda} \int_X e^{-\lambda A(x)} \, dx } \\ \\
&=& \displaystyle{ -\frac{1}{Z} \frac{dZ}{d \lambda} } \\ \\
&=& \displaystyle{ - \frac{d}{d \lambda} \ln Z .} \end{array} \]
This gives:
\[ \begin{array}{ccl} Q &=& \lambda \,\langle A \rangle + \ln Z \\ \\ 
&=& \displaystyle{ - \lambda \,\frac{d \ln Z }{d  \lambda}  + \ln Z. } \end{array} \]
The following chart shows where our analogy stands now.

\vskip 1em
\begin{center}
\renewcommand{\arraystretch}{1.4}
\begin{tabular}{|c|c|} \hline 
\textbf{Statistical Mechanics} & \textbf{Quantum Mechanics} \\
\hline
states: $ x \in X$ & histories: $ x \in X$ \\
\hline
probabilities: $ p\maps X \to [0,\infty)$ & amplitudes: $ a\maps X \to \C $ \\
\hline
energy: $ E\maps X \to \R$ & action: $ A\maps X \to \R$ \\
\hline
temperature: $ T$ & Planck's constant times $ i$: $ i \hbar$ \\
\hline
coolness: $ \beta = 1/T$ & classicality: $ \lambda = 1/i \hbar$ \\
\hline
partition function: $ Z = \int_X e^{-\beta E(x)} \, dx$ & partition function: $ Z = \int_X e^{-\lambda A(x)} \, dx$ \\ 
\hline
Boltzmann distribution: $ p(x) = e^{-\beta E(x)}/Z$ & Feynman sum over histories: $ a(x) = e^{-\lambda A(x)}/Z$ \\
\hline
entropy: $ S = - \int_X p(x) \ln p(x) \, dx$ & quantropy: $ Q = - \int_X a(x) \ln a(x) \, dx$ \\  
\hline
expected energy: $ \langle E \rangle = \int_X p(x) E(x) \, dx$ & expected action: $ \langle A \rangle = \int_X a(x) A(x) \, dx $ \\  
\hline
free energy: $ F = \langle E \rangle - TS$ & free action: $ \Phi = \langle A \rangle - i \hbar Q$ \\
\hline
$ \langle E \rangle = - \frac{d}{d \beta} \ln Z$ & $ \langle A \rangle = - \frac{d}{d \lambda} \ln Z$ \\
\hline
$ F = -\frac{1}{\beta} \ln Z$ & $ \Phi = -\frac{1}{\lambda} \ln Z$ \\
\hline
$ S =  \ln Z - \beta \,\frac{d}{d \beta}\ln Z $ & $ Q = \ln Z - \lambda \,\frac{d }{d \lambda}\ln Z $ \\  \hline
principle of maximum entropy & principle of stationary quantropy \\ \hline
principle of minimum energy & principle of stationary action \\ 
(in $T \to 0$ limit)   & (in $\hbar \to 0$ limit) \\ 
\hline
\end{tabular}
\end{center}

\vskip 2em


\section{The quantropy of a free particle}
\label{free_particle}

Let us illustrate these ideas with an example: a free particle.
Suppose we have a free particle on a line tracing out some path as time goes by:
\[ q\maps [0,T] \to \R \]
Then its action is just the time integral of its kinetic energy:
\[ \displaystyle{ A(q) = \int_0^T \frac{mv(t)^2}{2} \; dt }. \]
where  $v(t) = \dot q(t)$.  The partition function is then
\[ Z = \displaystyle{\int e^{i A(q) / \hbar} \; Dq } \]
where we integrate an exponential involving the action over the space of all paths.   

Unfortunately, the space of all paths is infinite-dimensional, so $Dq$ is ill-defined:
there is no `Lebesgue measure' on an infinite-dimensional vector space.  
So, we start by treating time as discrete---a trick going back to Feynman's original
work \cite{FeynmanHibbs}.   We consider $n$ time intervals of length $ \Delta t.$    
We say the position of our particle at the $ i$th time step is $ q_i \in \R$,
and require that the particle keeps a constant velocity $v_i$ between the $(i-1)$st 
and $ i$th time steps:
\[ \displaystyle{ v_i = \frac{q_i - q_{i-1}}{\Delta t} }. \]
Then the action, defined as an integral, reduces to a finite sum:
\[ \displaystyle{ A(q) = \sum_{i = 1}^n \frac{mv_i^2}{2} \; \Delta t } .\]
We consider histories of the particle where its initial position is $q_0 = 0$,
but its final position $ q_n$ is arbitrary.  If we do not `nail down' the particle at some particular time in this way, our path integrals will diverge.   So, our space of histories is
\[ X = \R^n \]
and we are ready to apply the formulas in the previous section.

We start with the partition function.  Naively, it is
\[ \displaystyle{  Z = \int_X e^{- \lambda A(q)} Dq } \]
where 
\[ \displaystyle{ \lambda = \frac{1}{i \hbar} } .\]
But this means nothing until we define the measure $Dq$.  Since the space of histories is just $ \R^n$ with coordinates $ q_1, \dots, q_n,$ an obvious guess for a measure would be
\[ Dq = dq_1 \cdots dq_n  . \]
However, the partition function should be dimensionless.  The quantity $ \lambda A(q)$ 
and its exponential are dimensionless, so the measure had better be dimensionless too. But $ dq_1 \cdots dq_n$ has units of $\textrm{length}^n$.   So to make the measure dimensionless, we introduce a length scale, $ \Delta x,$ and use the measure
\[ Dq = \displaystyle{ \frac{1}{(\Delta x)^n} \, dq_1 \cdots dq_n }.  \]
 It should be emphasized that despite the notation $ \Delta x,$ space is not discretized, just time.    This length scale $ \Delta x$ is introduced merely  order to make the measure on the space of histories  dimensionless.

Now let us compute the partition function.  For starters, we have
\[ \begin{array}{ccl} Z &=& \displaystyle{ \int_X e^{-\lambda A(q)} \; Dq } \\  \\ &=& \displaystyle{  \frac{1}{(\Delta x)^n} \int e^{-\lambda m \Delta t 
\sum_{i=1}^n  v_i^2 /2} \; dq_1 \cdots dq_n .} \end{array} \]
Since $q_0$ is fixed, we can express the positions $q_1, \dots , q_n$ in terms of the velocities $v_1, \dots v_n$.   Since
\[ dq_1 \cdots dq_n = (\Delta t)^n \; dv_1 \cdots dv_n \]
this change of variables gives
\[ Z = \displaystyle{\left(\frac{\Delta t}{\Delta x}\right)^n \int e^{-\lambda m \Delta t  \sum_{i=1}^n  v_i^2 /2} \; dv_1 \cdots dv_n }. \]
But this $n$-tuple integral is really just a product of $n$ integrals over one variable, all of which are equal.  So, we get some integral to the $n$th power:
\[ Z =  \displaystyle{ \left(\frac{\Delta t}{\Delta x}  \int_{-\infty}^\infty e^{-\lambda  m \Delta t \, v^2 /2} \; dv \right)^n }. \]

Now, when $\alpha$ is positive we have
\[  \displaystyle{ \int_{-\infty}^\infty e^{-x^2/2\alpha} \; d x = \sqrt{2 \pi \alpha}  } \]
but we will apply this formula to compute the partition function, where the constant playing the role of $\alpha$ is imaginary.  This makes some mathematicians nervous, because when $\alpha$ is imaginary, the function being
integrated is no longer Lebesgue integrable.  However, when $\alpha$ is imaginary, we get the same answer if we impose a cutoff and then let it go to infinity:
\[ \displaystyle{ \lim_{M \to + \infty} \int_{-M}^M e^{-x^2 / 2 \alpha} \, dx = \sqrt{2 \pi \alpha} } \]  
or damp the oscillations and then let the amount of damping go to zero:
\[ \displaystyle{ \lim_{\epsilon \downarrow 0} \int_{-\infty}^\infty e^{ -x^2 / 2\alpha \; - \;\epsilon x^2} \, dx = \sqrt{2 \pi \alpha} }. \]
So we shall proceed unabashed, and claim
\[ Z = \displaystyle{ \left( \frac{\Delta t}{\Delta x} \sqrt{ \frac{2 \pi}{\lambda m \, \Delta t}} \right)^n =  \left(\frac{2 \pi \Delta t}{\lambda 
m \, (\Delta x)^2}\right)^{n/2}  } .\]

Given this formula for the partition function, we can compute everything we care about:
the expected action, free action and quantropy.  Let us start with the expected action:
\[ \begin{array}{ccl} 
\langle A \rangle 
&=& 
\displaystyle{ -\frac{d}{d \lambda} \ln Z } \\  \\
&=& 
\displaystyle{ -\frac{n}{2}  \frac{d}{d \lambda}  \ln \left(\frac{2 \pi \Delta t}{\lambda m \, (\Delta x)^2}\right) } \\  \\ 
&=& 
\displaystyle{ \frac{n}{2}  \frac{d}{d \lambda} \left( \ln \lambda - \ln \left(\frac{2 \pi \Delta t}{m \, (\Delta x)^2}\right) \right) } \\   \\
&=& 
\displaystyle{ \frac{n}{2} \; \frac{1}{\lambda} } \\  \\
&=& 
 \displaystyle{ n\;  \frac{i \hbar}{2} }
\end{array} \]

This formula says that the expected action of our freely moving quantum particle is proportional to $ n,$ the number of time steps.  Each 
time step contributes $ i \hbar / 2$ to the expected action.  The mass of the particle, the time step $ \Delta t,$ and the length scale $ \Delta 
x$ do not matter at all; they disappear when we take the derivative of the logarithm 
containing them.  Indeed, our action could be any function of this sort:
\[ A \maps \R^n \to \R \]
\[ \displaystyle{ A(x) = \sum_{i = 1}^n \frac{c_i x_i^2}{2} } \]
where $ c_i$ are positive numbers, and we would still get the same expected action:
\[ \langle A \rangle = \displaystyle{ n\; \frac{i \hbar}{2} } \]
And since we can diagonalize any positive definite quadratic form, we can state 
this fact more generally: whenever the action is a positive definite quadratic form on an 
$n$-dimensional vector space of histories, the expected action is $n$ times $ i 
\hbar / 2.$  For example, consider a free particle in 3-dimensional Euclidean space, and discretize time into $n$ steps as we have done here.  Then the action is a positive definite quadratic form on a $3n$-dimensional vector space, so the expected action is $ 3n$ times 
$ i \hbar / 2.$

We can try to intepret this as follows.  In the path integral approach to quantum mechanics, a system can trace out any history it wants.  If the space of histories is an $n$-dimensional vector space, it takes $n$ real numbers to determine a specific history.  Each number counts as one `decision'.  And in the situation we have described, where the action is a positive definite quadratic form, each decision contributes $ i \hbar / 2$ to the expected action.  

There are some questions worth answering:

\begin{enumerate}
\item {\bf Why is the expected action imaginary?} The action $ A$ is real.  How can its expected value be imaginary?  The reason is that we are not taking its expected value with respect to a probability measure, but instead, with respect to a complex-valued  measure.  Recall that
\[ \langle A \rangle = \displaystyle{  \frac{\int_X A(x) e^{-\lambda A(x)} \, dx }{\int_X e^{-\lambda A(x)} \, dx }} .\]
The action $ A$ is real, but $ \lambda = 1 / i \hbar$ is imaginary, so it is not surprising that this `expected value' is complex-valued.  

\vskip 1em
\item {\bf Why does the expected action diverge as \define{$n \to \infty $}?}
We have discretized time in our calculation. To take the continuum limit we must let $ n \to \infty$ while simultaneously letting $ \Delta t \to 0$ in such a way that $ n \Delta t$ stays constant.  Some quantities will converge when we take this limit, but the expected action will not: it will go to infinity.  What does this mean?

This phenomenon is similar to how the expected length of the path of a particle undergoing Brownian motion is infinite.  In fact the free quantum particle is just a Wick-rotated version of Brownian motion, where we replace time by imaginary time, so the analogy  is fairly close.  The action we are considering now is not exactly analogous to the arclength of a path:
\[ \displaystyle{ \int_0^T \left| \frac{d q}{d t} \right| \; dt } \]
Instead, it is proportional to this quadratic form:
\[  \displaystyle{ \int_0^T \left| \frac{d q}{d t} \right|^2 \; dt }. \]
However, both these quantities diverge when we discretize Brownian motion and then take the continuum limit.   The reason is that for Brownian motion, with probability one the path of the particle is nondifferentiable, with Hausdorff dimension $\gt 1$ \cite{MP}.  We cannot apply probability theory to the quantum situation, but we are seeing that the `typical' path of a quantum free particle has infinite expected action in
the continuum limit.

\vskip 1em
\item {\bf Why does the expected action of the free particle resemble the expected energy of an ideal gas?}  For a classical ideal gas with $n$ particles in 3d space, the expected energy is
\[   \langle E \rangle =  \frac{3}{2} n T \]
in units where Boltzmann's constant is 1.  For a free quantum particle in 3d space, with time discretized into $n$ steps, the expected action is 
\[   \langle A \rangle = \frac{3}{2} n i\hbar.  \]
Why are the answers so similar?  

The answers are similar because of the analogy we are discussing.  Just as the action of the free particle is a positive definite quadratic form on $\R^n$, so is the energy of the ideal gas.  Thus, computing the expected action of the free particle is just like computing
the expected energy of the ideal gas, after we make these replacements:
 \[ \begin{array}{rcl} E &\mapsto& A \\
 T &\mapsto& i \hbar. \end{array}\]
\end{enumerate}

The last remark also means that the formulas for the free action and quantropy of a quantum free particle will be analogous those for the free energy and entropy of a classical ideal gas, except missing the factor of 3 when we consider a particle on a line.  For the free particle on a line, we have seen that
\[   \ln Z = \displaystyle{ \frac{n}{2}  \ln \left(\frac{2 \pi \Delta t}{\lambda m \, (\Delta x)^2}\right) }. \]
Setting
\[          K = \displaystyle{ \frac{2 \pi \Delta t}{m \, (\Delta x)^2} } ,\]
we can write this more compactly as
\[   \ln Z = \displaystyle{ \frac{n}{2} \left( \ln K - \ln \lambda \right) }.\]
We thus obtain the following formula for the free action: 
\[   \begin{array}{ccl} \Phi &=& -\displaystyle{ \frac{1}{\lambda} \ln Z } \\  \\
&=&  \displaystyle{  \frac{1}{\lambda} \frac{n}{2}\left( \ln \lambda - \ln K \right)   }. 
\end{array}
\]
Note that the $\ln K$ term dropped out when we computed the expected action by differentiating $\ln Z$ with respect to $\lambda$, but it shows up in the free action.  
 
The presence of this $\ln K$ term is surprising, since the constant $K$ is
not part of the usual theory of a free quantum particle.  A completely
analogous surprise occurs when computing the partition function of a classical ideal gas.  
The usual textbook answer involves a term of type $\ln K$  where $K$ is proportional to the volume of the box containing the gas divided by the cube of the thermal de Broglie wavelength of the gas molecules \cite{Reif}.  Curiously, the latter quantity involves Planck's constant, despite the fact that we we are considering 
a \emph{classical} ideal gas!   Indeed, we are \emph{forced} to introduce a quantity with dimensions of action to make the partition function of the gas dimensionless, because 
the partition function is an integral of a dimensionless quantity over position-momentum pairs, and $dp dq$ has units of action.  Nothing within classical mechanics forces us to choose this quantity to be Planck's constant; any choice will do.  Changing our choice only changes the free energy by an additive constant.  Nonetheless, introducing Planck's constant has the advantage of removing this ambiguity in the free energy of the classical ideal gas, in a way which is retroactively justified by quantum mechanics.

Analogous remarks apply to the length scale $\Delta x$ in our computation of the free action of a quantum particle.   We introduced it only to make the partition function dimensionless.  It is mysterious, much as Planck's constant was mysterious when it first forced its way into thermodynamics.  We do not have a theory or experiment that chooses a favored value for this constant.  All we can say at present is that it appears naturally when we push the analogy between statistical mechanics and quantum mechanics to its logical conclusion---or, a skeptic might say, to its breaking point.

Finally, the quantropy of the free particle on a line is 
\[ \begin{array}{ccl} Q
&=& \displaystyle{ - \lambda \,\frac{d \ln Z}{d \lambda} + \ln Z } \\  \\
&=&  \displaystyle{ \frac{n}{2} \left( \ln K  - \ln \lambda + 1 \right) }.
\end{array} \]
Again, the answer depends on the constant $K$: if we do not choose 
a value for this constant, we only obtain the quantropy up to an additive constant.
An analogous problem arises for the entropy of a classical ideal gas: without
introducing Planck's constant, we can only compute this entropy up to an additive
constant.


\section{Conclusions}
\label{conclusions}

There are many questions left to tackle.  The biggest is: \emph{what is the meaning of quantropy?}  Unfortunately it seems hard to attack this directly.  It may be easier to work out more examples and develop more of an intuition for this concept.  There are, however, some related puzzles worth keeping in mind.

As emphasized by Lisi \cite{Lisi}, it is rather peculiar that in the path-integral
approach to quantum mechanics we normalize the complex numbers $a(x)$ associated to paths so that they integrate to 1:
\[             \int_X a(x) \, dx = 1 .\]
It clearly makes sense to normalize \emph{probabilities} so that they sum to 1.  However, starting from the wavefunction of a quantum system, we obtain probabilities only after taking the absolute value of the wavefunction and squaring it.  Thus, for wavefunctions we impose
\[              \int_X |\psi(x)|^2 \, dx = 1  \]
rather than
\[              \int_X \psi(x) \, dx = 1 . \]
For this reason Lisi calls the numbers $a(x)$ `complex probabilities' rather than amplitudes.  However, the meaning of complex probabilities remains mysterious,
and this is tied to the mysterious nature of quantropy.  Feynman's essay on the interpretation of negative probabilities could provide some useful clues \cite{Feynman}.

It is also worth keeping in mind another analogy: `coolness as imaginary
time'.  Here we treat $\beta$ as analogous to $i t/ \hbar$ rather than $1/i \hbar$. This is widely used to convert quantum mechanics problems into statistical mechanics problems by means of Wick rotation, which essentially means studying the
unitary group $\exp(-i t H /\hbar) $ by studying the semigroup $\exp(-\beta H)$ and then analytically continuing $\beta$ to imaginary values.   Wick rotation plays an important role in Hawking's computation of the entropy of a black hole, nicely summarized in his book with Penrose \cite{HawkingPenrose}.  The precise relation of this other analogy to the one explored here remains unclear, and is worth exploring.  Note that the quantum 
Hamiltonian $H$ shows up on \emph{both} sides of this other analogy.


\subsection*{Acknowledgments}

We thank Garrett Lisi, Joakim Munkhammar, and readers of the \emph{Azimuth} blog
for many helpful suggestions.  We thank the Centre for Quantum Technology and an FQXi minigrant for supporting this research.

\end{document}